\font\tenfrakturb=eufb10
\font\tenfraktur=eufm10
\font\tenmsbm=msbm10
\font\sevenfrakturb=eufb7
\font\sevenfraktur=eufm7
\font\sevenmsbm=msbm7
\font\fivefrakturb=eufb5
\font\fivefraktur=eufm5
\font\fivemsbm=msbm5
\newfam\bgothicfam
\newfam\gothicfam
\newfam\msbmfam
\textfont\bgothicfam = \tenfrakturb \scriptfont\bgothicfam=\sevenfrakturb
\scriptscriptfont\bgothicfam=\fivefrakturb
\textfont\gothicfam = \tenfraktur \scriptfont\gothicfam=\sevenfraktur
\scriptscriptfont\gothicfam=\fivefraktur
\textfont\msbmfam = \tenmsbm \scriptfont\msbmfam=\sevenmsbm
\scriptscriptfont\msbmfam=\fivemsbm

\def\Bbb{\tenmsbm\fam\msbmfam}

\catcode`@=11
\def\renewcounter#1{\@definecounter{#1}\@ifnextchar[{\@newctr{#1}}{}}

\documentstyle{elsart}
\begin{document}
\begin{frontmatter}
\title{ Estimates for parameters and characteristics of the confining 
SU(3)-gluonic field in the ground state of toponium: Relativistic and 
nonrelativistic approaches}
\author{Yu. P. Goncharov}
\address{Theoretical Group, Experimental Physics Department, State Polytechnical
         University, Sankt-Petersburg 195251, Russia}

\begin{abstract} 
The confinement mechanism earlier proposed by author is applied to describe 
the (possible) ground state of toponium $\eta_t$. For this aim the 
nonperturbative consistent approach is elaborated in both relativistic and 
nonrelativistic cases. The study entails estimates 
for parameters of the confining SU(3)-gluonic field in the above quarkonium, 
those estimates being also consistent with possible width of decay 
$\eta_t\to2\gamma$. 
The corresponding estimates of the gluon concentrations, electric and magnetic 
colour field strengths are also adduced for the 
mentioned field at the scales of toponium. 
\end{abstract}
\end{frontmatter}   
\section{Introduction}

In Refs. \cite{{Gon01},{Gon051},{Gon052}} for the Dirac-Yang-Mills 
system derived from 
QCD-Lagrangian there was found and explored an unique family of compatible 
nonperturbative solutions 
which could pretend to decsribing confinement of two quarks. 
Applications of the family to description of both the heavy quarkonia 
spectra \cite{Gon03} and a number of properties of pions, kaons and 
$\eta$-meson \cite{{Gon06},{Gon07a},{Gon07b}} showed that the confinement 
mechanism is qualitatively the same for both light mesons and heavy quarkonia. 
At the moment it can be decribed in the following form. 

Two main physical reasons for linear confinement in the mechanism under 
discussion are the following ones. The first one is that gluon exchange between 
quarks is realized with the propagator different from the photon-like one and 
existence and form of such a propagator is {\em direct} consequence of the unique confining 
nonperturbative solutions of the Yang-Mills equations 
\cite{{Gon051},{Gon052}}. The second reason is that, 
owing to the structure of mentioned propagator, quarks mainly emit and 
interchange with the soft gluons so the gluon condensate (a classical gluon field) 
between quarks basically consists of soft gluons (for more details 
see Refs. \cite{{Gon051},{Gon052}}) but, because of that any gluon also emits 
gluons (still softer), the corresponding gluon concentrations 
rapidly become huge and form the linear confining magnetic colour field of 
enormous strengths which leads to confinement of quarks. This is by virtue of 
the fact that just magnetic part of the mentioned propagator is responsible 
for larger portion of gluon concentrations at large distances since the 
magnetic part has stronger infrared singularities than the electric one. 
Under the circumstances 
physically nonlinearity of the Yang-Mills equations effectively vanishes so the 
latter possess the unique nonperturbative confining solutions of the 
Abelian-like form (with the values in Cartan subalgebra of SU(3)-Lie algebra) 
\cite{{Gon051},{Gon052}} that describe 
the gluon condensate under consideration. Moreover, since the overwhelming majority 
of gluons is soft they cannot leave hadron (meson) until some gluon obtains 
additional energy (due to an external reason) to rush out. So we deal with 
confinement of gluons as well. 

The approach under discussion equips us with the explicit wave functions 
for every two quarks (meson or quarkonium). The wave functions are parametrized by a set of 
real constants $a_j, b_j, B_j$ describing the mentioned 
{\em nonperturbative} confining gluon field (the gluon condensate) and they  
are {\em nonperturbative} modulo square integrable 
solutions of the Dirac equation in the above confining SU(3)-field and also  
depend on $\mu_0$, the reduced
mass of the current masses of quarks forming meson. It is clear that under the 
given approach just constants $a_j, b_j, B_j,\mu_0$ determine all properties 
of any meson (quarkonium), i. e.,  the approach directly appeals to quark 
and gluonic degrees of freedom as should be according to the first principles 
of QCD. Also it is clear that the mentioned constants should be extracted from 
experimental data. 

Such a program has been to a certain extent advanced in 
Refs. \cite{{Gon03},{Gon06},{Gon07a},{Gon07b}}. The aim of the present paper is 
to return to heavy quarkonia physics to obtain estimates for $a_j, b_j, B_j$ in  
the possible ground state of the heaviest quarkonium -- toponium (in what 
follows we denote it as $\eta_t$), where so far 
little is still known about experimental spectroscopy of the system. 

Under the situation a certain motivation of studying 
toponium is adduced in Section 2. Section 3 contains main relations underlying 
both relativistic and nonrelativistic descriptions of any quarkonia in our 
approach. Section 4 is 
devoted to computing electric form factor, the root-mean-square radius $<r>$ 
and magnetic moment of the quarkonium under consideration in an explicit 
analytic form. Section 5 gives an independent estimate for $<r>$ which is used 
in Section 6 for obtaining estimates for parameters of the confining 
SU(3)-gluonic field for the toponium ground state $\eta_t$ in both relativistic 
and nonrelativistic cases. Further in Section 7 we show that estimates of 
Section 6 can also be consistent with possible width of two-photon decay 
$\eta_t\to2\gamma$. Section 8 employs the obtained parameters of SU(3)-gluonic 
field to get the corresponding estimates for such characteristics of the 
mentioned field as gluon concentrations, electric and magnetic colour field 
strengths at the scales of $\eta_t$ while Section 9 is devoted to discussion 
and concluding remarks. 
                      
At last Appendices $A$ and $B$ contain the detailed description 
of main building blocks for meson wave functions in the approach under 
discussion, respectively: eigenspinors of the Euclidean Dirac operator on 
two-sphere ${\Bbb S}^2$ and radial parts for the modulo square integrable 
solutions of Dirac equation in the confining SU(3)-Yang-Mills field while 
Appendix $C$ supplements Appendices $A$ and $B$ with proof of the fact that 
the so-called nonrelativistic confining potentials do not obey the Maxwell or 
SU(3)-Yang--Mills equations. 
 
Further we shall deal with the metric of
the flat Minkowski spacetime $M$ that
we write down (using the ordinary set of local spherical coordinates
$r,\vartheta,\varphi$ for the spatial part) in the form
$$ds^2=g_{\mu\nu}dx^\mu\otimes dx^\nu\equiv
dt^2-dr^2-r^2(d\vartheta^2+\sin^2\vartheta d\varphi^2)\>. \eqno(1)$$
Besides, we have $|\delta|=|\det(g_{\mu\nu})|=(r^2\sin\vartheta)^2$
and $0\leq r<\infty$, $0\leq\vartheta<\pi$,
$0\leq\varphi<2\pi$.

Throughout the paper we employ the Heaviside-Lorentz system of units 
with $\hbar=c=1$, unless explicitly stated otherwise, so the gauge coupling 
constant $g$ and the strong coupling constant ${\alpha_s}$ are connected by 
relation $g^2/(4\pi)=\alpha_s$. 
In what follows we shall denote $L_2(F)$ the set of the modulo square integrable
complex functions on any manifold $F$ furnished with an integration measure, 
then $L^n_2(F)$ will be the $n$-fold direct product of $L_2(F)$
endowed with the obvious scalar product while $\dag$ and $\ast$ stand, 
respectively, for Hermitian and complex conjugation. Our choice of Dirac 
$\gamma$-matrices conforms to the so-called standard representation and is 
the same as in Ref. \cite{Gon06}. At last $\otimes$ means 
tensorial product of matrices and $I_n$ is the unit $n\times n$ matrix so that, 
e.g., we have 
$$I_3\otimes\gamma^\mu=
\pmatrix{\gamma^\mu&0&0\cr 0&\gamma^\mu&0\cr 0&0&\gamma^\mu\cr}$$ 
for any Dirac $\gamma$-matrix $\gamma^\mu$ and so forth. 

When calculating we apply the 
relations $1\ {\rm GeV^{-1}}\approx0.1973269679\ {\rm fm}\>$,
$1\ {\rm s^{-1}}\approx0.658211915\times10^{-24}\ {\rm GeV}\>$, 
$1\ {\rm V/m}\approx0.2309956375\times 10^{-23}\ {\rm GeV}^2$, 
$1\ {\rm T}=4\pi\times10^{-7} {\rm H/m}\times1\ {\rm A/m}
\approx0.6925075988\times 10^{-15}\ {\rm GeV}^2 $. 

Finally, for the necessary estimates we shall employ the $T_{00}$-component 
(volumetric energy density ) of the energy-momentum tensor for a 
SU(3)-Yang-Mills field which should be written in the chosen system of units 
in the form
$$T_{\mu\nu}=-F^a_{\mu\alpha}\,F^a_{\nu\beta}\,g^{\alpha\beta}+
{1\over4}F^a_{\beta\gamma}\,F^a_{\alpha\delta}g^{\alpha\beta}g^{\gamma\delta}
g_{\mu\nu}\>. \eqno(2) $$

\section{Motivation}
Though theoretical toponium physics has been developing already during a long 
time (see early reviews of Refs. \cite{top} and references therein) so far 
little is still known 
about experimental spectroscopy of the system, as we have mentioned in 
Section 1. 
According to standard model (SM) with three generations (see, e.g., 
Ref. \cite{pdg}) main properties of top-quark are determined by summand in 
SM-Lagrangian of the form $-g/(2\sqrt{2})[\bar{t}\gamma_\mu(1+
\gamma_5)V_{tb}\,b]W^\mu$, where the gauge coupling constant $g$ is connected 
with the Fermi coupling constant $G_F$ as $g^2\sqrt{2}=8G_Fm^2_W$ 
($m_W\approx80.403$ Gev is mass of $W$-boson) while 
$V_{tb}$ is the corresponding element of the Cabibbo--Kobayashi--Maskawa 
mixing matrix. In second order in $g$ this yields the decay width 
$\Gamma(t\to W\,b)\sim G_F|V_{tb}|^2m^3_t$ so that $\Gamma$ proves to be 
between (1.0 -- 1.6) GeV depending on top-quark mass $m_t$ \cite{pdg}. With its 
respectively short lifetime of order $0.5\times10^{-24}$ s, the top quark is 
expected to decay before top-flavored hadrons or $\bar{t}t$-quarkonium bound 
states can form although as far back as in Ref. \cite{Fad87} possibilities for  
formation of toponium at $m_t\sim170$ GeV were discussed (see also 
Refs. \cite{Kho98} and references therein). 

  The above estimates, however, suppose $|V_{tb}|$ to be of order 1. But it is 
known that mixings and the number of fermion generations are not fixed by SM. 
Under the circumstances possible existence of extra SM families may 
sufficiently decrease $|V_{tb}|$ so that toponium can be formed \cite{Cak04}.  
Under this situation, one of the main signals for detecting toponium should 
be decay $\eta_t\to2\gamma$ and enough number of those decays might be 
observable even at LHC and its further upgrades and, of course, at possible 
future collider VLHC (for more details see Ref. \cite{Cak04}). For the sake of 
justice we should, however, note that present measurements still indicate 
that $|V_{tb}|>$0.78 \cite{pdg}. 

Also within the SM based on the two Higgs doublets formation and observation 
of heavy quarkonia (including toponium) might be lightened \cite{Koz03}. Thus, 
on the whole, the question of existence and observability for toponium remains 
open and, probably, possible discovery of toponium at future colliders might 
provide a connection to new or unexpected physics.

 Specifically, therefore, there is an certain interest of trying to explore 
the toponium already now from the point of view of the above confinement 
mechanism which directly appeals to quark and gluonic degrees of freedom as 
should be according to the first principles 
of QCD. Under the circumstances we shall use some previous theoretical 
estimates for toponium obtained from other considerations (see, e. g., Refs. 
\cite{{Yn01},{Ki03}} and references therein).

\section{Main relations}

As was mentioned above, our considerations shall be based on the unique family 
of compatible nonperturbative solutions for 
the Dirac-Yang-Mills system (derived from QCD-Lagrangian) studied in details 
in Refs. \cite{{Gon01},{Gon051},{Gon052}}.  Referring for more details to those 
references, let us briefly describe and specify only the relations necessary to 
us in the present paper. 

One part of the mentioned family is presented by the unique nonperturbative 
confining solution of the Yang-Mills 
equations for $A=A_\mu dx^\mu=A^a_\mu \lambda_adx^\mu$ ($\lambda_a$ are the 
known Gell-Mann matrices, $\mu=t,r,\vartheta,\varphi$, $a=1,...,8$) and looks 
as follows 
$$ {\cal A}_{1t}\equiv A^3_t+\frac{1}{\sqrt{3}}A^8_t =-\frac{a_1}{r}+A_1 \>,
{\cal A}_{2t}\equiv -A^3_t+\frac{1}{\sqrt{3}}A^8_t=-\frac{a_2}{r}+A_2\>,$$
$${\cal A}_{3t}\equiv-\frac{2}{\sqrt{3}}A^8_t=\frac{a_1+a_2}{r}-(A_1+A_2)\>, $$
$$ {\cal A}_{1\varphi}\equiv A^3_\varphi+\frac{1}{\sqrt{3}}A^8_\varphi=
b_1r+B_1 \>,
{\cal A}_{2\varphi}\equiv -A^3_\varphi+\frac{1}{\sqrt{3}}A^8_\varphi=
b_2r+B_2\>,$$
$${\cal A}_{3\varphi}\equiv-\frac{2}{\sqrt{3}}A^8_\varphi=
-(b_1+b_2)r-(B_1+B_2)\> \eqno(3)$$
with the real constants $a_j, A_j, b_j, B_j$ parametrizing the family. 
As has been repeatedly explained in 
Refs. \cite{{Gon051},{Gon052},{Gon03},{Gon06}}, parameters $A_{1,2}$ of 
solution (3) are inessential for physics in question and we can 
consider $A_1=A_2=0$. Obviously we have 
$\sum_{j=1}^{3}{\cal A}_{jt}=\sum_{j=1}^{3}{\cal A}_{j\varphi}=0$ which 
reflects the fact that for any matrix 
${\cal T}$ from SU(3)-Lie algebra we have ${\rm Tr}\,{\cal T}=0$. 
Also, as has been repeatedly discussed by us earlier (see, e. g., 
Refs. \cite{{Gon051},{Gon052}}), from the above form it is clear that 
the solution (3) is a configuration describing the electric Coulomb-like colour 
field (components $A^{3,8}_t$) and the magnetic colour field linear in $r$ 
(components $A^{3,8}_\varphi$) and we wrote down
the solution (3) in the combinations that are just 
needed further to insert into the Dirac equation (4). 

Another part of the family is given by the unique nonperturbative modulo 
square integrable solutions of the Dirac equation in the confining 
SU(3)-field of (3) $\Psi=(\Psi_1, \Psi_2, \Psi_3)$ 
with the four-dimensional Dirac spinors 
$\Psi_j$ representing the $j$th colour component of the meson, 
so $\Psi$ may describe relative motion (relativistic bound states) of two quarks 
in mesons and the mentioned Dirac equation looks as follows 
$$i\partial_t\Psi\equiv  
i\pmatrix{\partial_t\Psi_1\cr \partial_t\Psi_2\cr \partial_t\Psi_3\cr}=
H\Psi\equiv\pmatrix{H_1&0&0\cr 0&H_2&0\cr 0&0&H_3\cr}
\pmatrix{\Psi_1\cr\Psi_2\cr\Psi_3\cr}=
\pmatrix{H_1\Psi_1\cr H_2\Psi_2\cr H_3\Psi_3\cr}
                   \,,\eqno(4)$$
where Hamiltonian $H_j$ is 
$$H_j=\gamma^0\left[\mu_0-i\gamma^1\partial_r-i\gamma^2\frac{1}{r}
\left(\partial_\vartheta+\frac{1}{2}\gamma^1\gamma^2\right)-
i\gamma^3\frac{1}{r\sin{\vartheta}}
\left(\partial_\varphi+\frac{1}{2}\sin{\vartheta}\gamma^1\gamma^3
+\frac{1}{2}\cos{\vartheta}\gamma^2\gamma^3\right)\right]$$
$$-g\gamma^0\left(\gamma^0{\cal A}_{jt}+\gamma^3\frac{1}{r\sin{\vartheta}}
{\cal A}_{j\varphi}\right) \eqno(5)  $$                           
with the gauge coupling constant $g$ while $\mu_0$ is a mass parameter and one 
should consider it to be the reduced mass which is equal, {\it e. g.}, for 
quarkonia, to half the current mass of quarks forming a quarkonium.

Then the unique nonperturbative modulo square integrable solutions of (4) 
are (with Pauli matrix $\sigma_1$)  
$$\Psi_j=e^{-i\omega_j t}\psi_j\equiv 
e^{-i\omega_j t}r^{-1}\pmatrix{F_{j1}(r)\Phi_j(\vartheta,\varphi)\cr\
F_{j2}(r)\sigma_1\Phi_j(\vartheta,\varphi)}\>,j=1,2,3\eqno(6)$$
with the 2D eigenspinor $\Phi_j=\pmatrix{\Phi_{j1}\cr\Phi_{j2}}$ of the
Euclidean Dirac operator ${\cal D}_0$ on the unit sphere ${\Bbb S}^2$, while 
the coordinate $r$ stands for the distance between quarks. The explicit form of 
$\Phi_j$ is discussed in Appendix $A$. We can call the quantity $\omega_j$ 
relative energy of $j$th colour component of meson (while $\psi_j$ is wave 
function of a stationary state for $j$th colour component) but we can see that 
if we want to interpret (4) as equation for eigenvalues of the relative 
motion energy, i. e.,  to rewrite it in the form $H\psi=\omega\psi$ with 
$\psi=(\psi_1, \psi_2, \psi_3)$ then we should put $\omega=\omega_j$ for 
any $j$ so that $H_j\psi_j=\omega_j\psi_j=\omega\psi_j$. Under this situation, 
if a meson is composed of quarks $q_{1,2}$ with different flavours then 
the energy spectrum of the meson will be given 
by $\epsilon=m_{q_1}+m_{q_2}+\omega$ with the current quark masses $m_{q_k}$ (
rest energies) of the corresponding quarks. On the other hand for 
determination of $\omega_j$ the following quadratic equation can be obtained 
\cite{{Gon01},{Gon051},{Gon052}}
$$[g^2a_j^2+(n_j+\alpha_j)^2]\omega_j^2-
2(\lambda_j-gB_j)g^2a_jb_j\,\omega_j+
[(\lambda_j-gB_j)^2-(n_j+\alpha_j)^2]g^2b_j^2-
\mu_0^2(n_j+\alpha_j)^2=0\>,  \eqno(7)   $$
that yields (at $g\ne0$) 
$$\omega_j=\omega_j(n_j,l_j,\lambda_j)=$$ 
$$\frac{\Lambda_j g^2a_jb_j\pm(n_j+\alpha_j)
\sqrt{(n_j^2+2n_j\alpha_j+\Lambda_j^2)\mu_0^2+g^2b_j^2(n_j^2+2n_j\alpha_j)}}
{n_j^2+2n_j\alpha_j+\Lambda_j^2}\>, j=1,2,3\>,\eqno(8)$$

where $a_3=-(a_1+a_2)$, $b_3=-(b_1+b_2)$, $B_3=-(B_1+B_2)$, 
$\Lambda_j=\lambda_j-gB_j$, $\alpha_j=\sqrt{\Lambda_j^2-g^2a_j^2}$, 
$n_j=0,1,2,...$, while $\lambda_j=\pm(l_j+1)$ are
the eigenvalues of Euclidean Dirac operator ${\cal D}_0$ 
on unit sphere with $l_j=0,1,2,...$. It should be noted that in the  
papers \cite{{Gon01},{Gon051},{Gon052},{Gon03},{Gon06}} we used the ansatz (6) 
with the factor $e^{i\omega_j t}$ instead of $e^{-i\omega_j t}$ but then the 
Dirac equation (4) would look as $-i\partial_t\Psi= H\Psi$ and in equation (7) 
the second summand would have plus sign while the first summand in numerator 
of (8) would have minus sign. In the papers \cite{{Gon07a},{Gon07b}}   
we returned to the conventional form of 
writing Dirac equation and this slightly modified the equations (7)--(8). In 
the given paper we conform to the same prescription as in 
Refs. \cite{{Gon07a},{Gon07b}}. 

In line with the above we should have $\omega=\omega_1=\omega_2=\omega_3$ in 
energy spectrum $\epsilon=m_{q_1}+m_{q_2}+\omega$ for any meson (quarkonium) 
and this at once imposes two conditions on parameters $a_j,b_j,B_j$ when 
choosing some experimental value for $\epsilon$ at the given current quark 
masses $m_{q_1},m_{q_2}$. 

The general form of the radial parts of (6) is considered in Appendix $B$. 
Within the given paper we need only of the radial parts of (6) at $n_j=0$ 
(the ground state) that are [see $(B.5)$]  
$$F_{j1}=C_jP_jr^{\alpha_j}e^{-\beta_jr}\left(1-
\frac{gb_j}{\beta_j}\right), P_j=gb_j+\beta_j, $$
$$F_{j2}=iC_jQ_jr^{\alpha_j}e^{-\beta_jr}\left(1+
\frac{gb_j}{\beta_j}\right), Q_j=\mu_0-\omega_j\eqno(9)$$
with $\beta_j=\sqrt{\mu_0^2-\omega_j^2+g^2b_j^2}$ while $C_j$ is determined 
from the normalization condition
$\int_0^\infty(|F_{j1}|^2+|F_{j2}|^2)dr=\frac{1}{3}$. 
Consequently, we shall gain that $\Psi_j\in L_2^{4}({\Bbb R}^3)$ at any 
$t\in{\Bbb R}$ and, as a result,
the solutions of (6) may describe relativistic bound states (mesons) 
with the energy (mass) spectrum $\epsilon$.
\subsection{Nonrelativistic limit}
It is useful to specify the nonrelativistic limit (when 
$c\to\infty$) for spectrum (8). For that one should replace 
$g\to g/\sqrt{\hbar c}$, 
$a_j\to a_j/\sqrt{\hbar c}$, $b_j\to b_j\sqrt{\hbar c}$, 
$B_j\to B_j/\sqrt{\hbar c}$ and, expanding (8) in $z=1/c$, we shall get
$$\omega_j(n_j,l_j,\lambda_j)=$$
$$\pm\mu_0c^2\left[1\mp
\frac{g^2a_j^2}{2\hbar^2(n_j+|\lambda_j|)^2}z^2\right]
+\left[\frac{\lambda_j g^2a_jb_j}{\hbar(n_j+|\lambda_j|)^2}\,
\mp\mu_0\frac{g^3B_ja_j^2f(n_j,\lambda_j)}{\hbar^3(n_j+|\lambda_j|)^{7}}\right]
z\,+O(z^2)\>,\eqno(10)$$
where 
$f(n_j,\lambda_j)=4\lambda_jn_j(n_j^2+\lambda_j^2)+
\frac{|\lambda_j|}{\lambda_j}\left(n_j^{4}+6n_j^2\lambda_j^2+\lambda_j^4
\right)$. 

As is seen from (10), at $c\to\infty$ the contribution of linear magnetic 
colour field (parameters $b_j, B_j$) to spectrum really vanishes and spectrum 
in essence becomes purely nonrelativistic Coulomb one (modulo the rest energy). Also it is 
clear that when $n_j\to\infty$, $\omega_j\to\pm\sqrt{\mu_0^2+g^2b_j^2}$. 

We may seemingly use (8) with various combinations of signes ($\pm$) before 
second summand in numerators of (8) but, due to (10), it is 
reasonable to take all signs equal to plus which is our choice within the 
paper. Besides, 
as is not complicated to see, radial parts in nonrelativistic limit have 
the behaviour of form $F_{j1},F_{j2}\sim r^{l_j+1}$, which allows one to call 
quantum number $l_j$ angular momentum for $j$th colour component though angular 
momentum is not conserved in the field (3) \cite{{Gon01},{Gon052}}. So for 
meson (quarkonium) under consideration we should put all $l_j=0$. 

\subsection{Eigenspinors with $\lambda=\pm1$}
Finally it should be noted that spectrum (8) is degenerated owing to 
degeneracy of eigenvalues for the
Euclidean Dirac operator ${\cal D}_0$ on the unit sphere ${\Bbb S}^2$. Namely,  
each eigenvalue of ${\cal D}_0$ $\lambda =\pm(l+1), l=0,1,2...$, has 
multiplicity $2(l+1)$ so we has $2(l+1)$ eigenspinors orthogonal to each other. 
Ad referendum we need eigenspinors corresponding to $\lambda =\pm1$ ($l=0$) 
so here is their explicit form [see $(A.16)$] 
$$\lambda=-1: \Phi=\frac{C}{2}\pmatrix{e^{i\frac{\vartheta}{2}}
\cr e^{-i\frac{\vartheta}{2}}\cr}e^{i\varphi/2},\> {\rm or}\>\>
\Phi=\frac{C}{2}\pmatrix{e^{i\frac{\vartheta}{2}}\cr
-e^{-i\frac{\vartheta}{2}}\cr}e^{-i\varphi/2},$$
$$\lambda=1: \Phi=\frac{C}{2}\pmatrix{e^{-i\frac{\vartheta}{2}}\cr
e^{i\frac{\vartheta}{2}}\cr}e^{i\varphi/2}, \> {\rm or}\>\>
\Phi=\frac{C}{2}\pmatrix{-e^{-i\frac{\vartheta}{2}}\cr
e^{i\frac{\vartheta}{2}}\cr}e^{-i\varphi/2} 
\eqno(11) $$
with the coefficient $C=1/\sqrt{2\pi}$ (for more details see 
Appendix $A$).

\section{Electric form factor, the root-mean-square radius and magnetic moment}

As has been mentioned in Section 1, at present little is known about 
experimental spectroscopy of the toponium so we should choose a few quantities 
that are the most important from the physical point of view to characterize the 
toponium and then we should evaluate the given quantities within the framework 
of our approach. Under the circumstances let us settle on the ground 
state energy of toponium, the root-mean-square radius of it and magnetic 
moment. All three magnitudes are essentially nonperturbative ones and can be 
calculated only by nonperturbative techniques. 

Within the present paper we shall use relations (8) at $n_j=0=l_j$ so the ground 
state energy of toponium is given by $\epsilon=2m_t+\omega$ with 
$\omega=\omega_j(0,0,\lambda_j)$ for any $j=1,2,3$ whereas 
$$\omega=\frac{g^2a_1b_1}{\Lambda_1}+\frac{\alpha_1\mu_0}
{|\Lambda_1|}=\frac{g^2a_2b_2}{\Lambda_2}+\frac{\alpha_2\mu_0}
{|\Lambda_2|}=\frac{g^2a_3b_3}{\Lambda_3}+\frac{\alpha_3\mu_0}
{|\Lambda_3|}=\epsilon-2m_t
\>\eqno(12)$$
and, as a consequence, the corresponding toponium wave functions of (6) are 
represented by (9) and (11). As the concrete value of $\epsilon$ we shall take 
the one of Ref. \cite{Ki03} equal to 347.4 GeV. 
\subsection{Choice of t-quark mass and gauge coupling constant}
It is evident for employing the above relations we have to assign some values 
to $t$-quark mass and gauge coupling constant $g$. In accordance with 
Ref. \cite{pdg} we take $m_t= 173.25$ GeV at present.  
Under the circumstances, the reduced mass $\mu_0$ of (5) will take value 
$m_t/2$. As to 
gauge coupling constant $g=\sqrt{4\pi\alpha_s}$, it should be noted that 
recently some attempts have been made to generalize the standard formula
for $\alpha_s=\alpha_s(Q^2)=12\pi/[(33-2n_f)\ln{(Q^2/\Lambda^2)}]$ ($n_f$ is 
number of quark flavours) holding true at the momentum transfer 
$\sqrt{Q^2}\to\infty$ 
to the whole interval $0\le \sqrt{Q^2}\le\infty$. We shall employ one such a 
generalization used in Refs. \cite{De1}. It looks as follows 
($x=\sqrt{Q^2}$ in GeV) 
$$ \alpha(x)=\frac{12\pi}{(33-2n_f)}\frac{f_1(x)}{\ln{\frac{x^2+f_2(x)}
{\Lambda^2}}} 
\eqno(13) $$
with 
$$f_1(x)=
1+\left(\left(\frac{(1+x)(33-2n_f)}{12}\ln{\frac{m^2}{\Lambda^2}}-1
\right)^{-1}+0.6x^{1.3}\right)^{-1}\>,\>f_2(x)=m^2(1+2.8x^2)^{-2}\>,$$
wherefrom one can conclude that $\alpha_s\to \pi=3.1415...$ when $x\to 0$, 
i. e., $g\to{2\pi}=6.2831...$. We used (13) at $m=1$ GeV, $\Lambda=0.234$ GeV, 
$n_f=6$, $x=2m_t=346.50$ GeV to obtain $g=1.243528161$ necessary for 
our further computations at the mass scale of toponium. 

\subsection{Electric form factor}
For each meson (quarkonium) with the wave function $\Psi=(\Psi_j)$ of (6) we 
can define 
electromagnetic current $J^\mu=\overline{\Psi}(I_3\otimes\gamma^\mu)\Psi=
(\Psi^{\dag}\Psi,\Psi^{\dag}(I_3\otimes{\bf \alpha})\Psi)=(\rho,{\bf J})$, 
${\bf \alpha}=\gamma^0{\bf\gamma}$.  
Electric form factor $f(K)$ is the Fourier transform of $\rho$
$$ f(K)= \int\Psi^{\dag}\Psi e^{-i{\bf K}{\bf r}}d^3x=\sum\limits_{j=1}^3
\int\Psi_j^{\dag}\Psi_j e^{-i{\bf K}{\bf r}}d^3x =\sum\limits_{j=1}^3f_j(K)=$$ 
$$\sum\limits_{j=1}^3
\int (|F_{j1}|^2+|F_{j2}|^2)\Phi_j^{\dag}\Phi_j
\frac{e^{-i{\bf K}{\bf r}}}{r^2}d^3x,\>
d^3x=r^2\sin{\vartheta}dr d\vartheta d\varphi\eqno(14)$$
with the momentum transfer $K$. At $n_j=0=l_j$, as is easily seen, for any  
spinor of (11) we have $\Phi_j^{\dag}\Phi_j=1/(4\pi)$, so the integrand in 
(14) does not depend on $\varphi$ and we can consider vector ${\bf K}$ to be 
directed along z-axis. Then ${\bf Kr}=Kr\cos{\vartheta}$ and with the help of 
(9) and relations (see Ref. \cite{PBM1}): $\int_0^\infty 
r^{\alpha-1}e^{-pr}dr=
\Gamma(\alpha)p^{-\alpha}$, Re $\alpha,p >0$, 
$\int_0^\infty r^{\alpha-1}e^{-pr}\pmatrix{\sin{(Kr)}\cr\cos{(Kr)}\cr}dr=
\Gamma(\alpha)(K^2+p^2)^{-\alpha/2}
\pmatrix{\sin{(\alpha\arctan{(K/p))}}\cr\cos{(\alpha\arctan{(K/p))}}\cr}$, 
Re $\alpha >-1$, 
Re $p > |{\rm Im}\, K|$, $\Gamma(\alpha+1)=\alpha\Gamma(\alpha)$, 
$\int_0^\pi e^{-iKr\cos{\vartheta}}\sin{\vartheta}d\vartheta=2\sin{(Kr)}/(Kr)$, 
we shall obtain 
$$ f(K)=\sum\limits_{j=1}^3f_j(K)=
\sum\limits_{j=1}^3\frac{(2\beta_j)^{2\alpha_j+1}}{6\alpha_j}\cdot
\frac{\sin{[2\alpha_j\arctan{(K/(2\beta_j))]}}}{K(K^2+4\beta_j^2)^{\alpha_j}}$$
$$=\sum\limits_{j=1}^3\left(\frac{1}{3}-\frac{2\alpha^2_j+3\alpha_j+1}
{6\beta_j^2}\cdot \frac{K^2}{6}\right)+O(K^4), \eqno(15)$$
wherefrom it is clear that $f(K)$ is a function of $K^2$, as should be, and 
we can determine the root-mean-square radius of meson (quarkonium) in the form 
$$<r>=\sqrt{\sum\limits_{j=1}^3\frac{2\alpha^2_j+3\alpha_j+1}
{6\beta_j^2}}.\eqno(16)$$
When calculating (15) also the fact was used that by virtue of the 
normalization condition for wave 
functions we have $C_j^2[P_j^2(1-gb_j/\beta_j)^2+Q_j^2(1+gb_j/\beta_j)^2]=
(2\beta_j)^{2\alpha_j+1}/[3\Gamma(2\alpha_j+1)]$.

It is clear, we can directly calculate $<r>$ in accordance with the standard 
quantum mechanics rules as $<r>=\sqrt{\int r^2\Psi^{\dag}\Psi d^3x}=
\sqrt{\sum\limits_{j=1}^3\int r^2\Psi^{\dag}_j\Psi_j d^3x}$ and the 
result will be the same as in (16). So we should not call $<r>$ of (16) 
the {\em charge} radius of meson (quarkonium)-- it is just the radius of meson 
(quarkonium) determined 
by the wave functions of (6) (at $n_j=0=l_j$) with respect to strong 
interaction, i.e., radius of confinement.  
Now we should notice the expression (15) to depend on 3-vector ${\bf K}$. To 
rewrite it in the form holding true for any 4-vector $Q$, let us remind that 
according to general considerations (see, e.g., Ref. \cite{LL1}) the relation 
(15) corresponds to the so-called Breit frame where $Q^2=-K^2$ [when fixing metric 
by (1)] so it is 
not complicated to rewrite (15) for arbitrary $Q$ in the form 
$$ f(Q^2)=\sum\limits_{j=1}^3f_j(Q^2)=
\sum\limits_{j=1}^3\frac{(2\beta_j)^{2\alpha_j+1}}{6\alpha_j}\cdot
\frac{\sin{[2\alpha_j\arctan{(\sqrt{|Q^2|}/(2\beta_j))]}}}
{\sqrt{|Q^2|}(4\beta_j^2-Q^2)^{\alpha_j}}\> \eqno(17) $$
which passes on to (15) in the Breit frame. 

\subsection{Magnetic moment}
We can define the volumetric magnetic moment density by 
${\bf m}=q({\bf r}\times {\bf J})/2=q[(yJ_z-zJ_y){\bf i}+
(zJ_x-xJ_z){\bf j}+(xJ_y-yJ_x){\bf k}]/2$ with the meson charge $q$ and 
${\bf J}=\Psi^{\dag}(I_3\otimes{\bf \alpha})\Psi$. Using (6) we have in the 
explicit form 
$$J_x=\sum\limits_{j=1}^3
(F^\ast_{j1}F_{j2}+F^\ast_{j2}F_{j1})\frac{\Phi_j^{\dag}\Phi_j}
{r^2},\> 
J_y=\sum\limits_{j=1}^3
(F^\ast_{j1}F_{j2}-F^\ast_{j2}F_{j1})
\frac{\Phi_j^{\dag}\sigma_2\sigma_1\Phi_j}{r^2},\>$$
$$J_z=\sum\limits_{j=1}^3
(F^\ast_{j1}F_{j2}-F^\ast_{j2}F_{j1})
\frac{\Phi_j^{\dag}\sigma_3\sigma_1\Phi_j}{r^2}  \eqno(18)$$
with Pauli matrices $\sigma_{1,2,3}$.
Magnetic moment of meson (quarkonium) is ${\bf M}=\int_V {\bf m}d^3x$, where 
$V$ is volume 
of meson (quarkonium) (the ball of radius $<r>$). Then at $n_j=l_j=0$, as is seen from (9), 
(11), $F^\ast_{j1}=F_{j1},F^\ast_{j2}=-F_{j2}$, 
$\Phi_j^{\dag}\sigma_2\sigma_1\Phi_j=0 $ for any spinor of (11) which entails 
$J_x=J_y=0$, i.e., $m_z=0$ while $\int_V m_{x,y}d^3x=0$ because of 
turning the integral over $\varphi$ to zero, which is easy to check.
As a result, magnetic moments of mesons (quarkonia) with the 
wave functions of (6) (at $l_j=0$) are equal to zero, as should be according 
to experimental data \cite{pdg}. 

Though we can also evaluate magnetic form factor $F(Q^2)$ of meson (quarkonium) 
as a function of $Q^2$ (see Refs. \cite{{Gon07a},{Gon07b}}) but the latter 
will not be used in the given paper so we shall not dwell upon it. 

\section{An estimate of $<r>$ from leptonic width}
The question now is how to estimate $<r>$ independently to then calculate it 
within framework of our approach. For this aim we shall employ the possible 
width of leptonic decay $V\to e^+e^-$ which is approximately equal 
to $\Gamma_1\approx13$ keV according to Ref. \cite{Yn01} and $V$ stands for 
the toponium state analogous to $J/\psi$ state in charmonium. Under this 
situation one can use a variant of formulas originating from Ref. \cite{Van67}. Such 
formulas are often employed in the heavy quarkonia physics (see, e. g., 
Ref. \cite{Bra05}). In their turn they are actually based on the standard 
expression from the elementary kinetic theory of gases (see, e. g., 
Ref. \cite{{Sav89}}) for the number $\nu$ of collisions of a molecule per unit 
time
$$ \nu=\sqrt{2}\sigma <v>n\>,                  \eqno(19)$$  
where $\sigma$ is an effective cross section for molecules, $<v>$ is a mean 
molecular velocity, $n$ is the concentration of molecules. 
If replacing $\nu\to\Gamma_1$ we may fit (19) to estimate the leptonic width 
$\Gamma_1=\Gamma_1(V\to e^+e^-)$ when interpreting $\sigma$ as the cross 
section of creation of $e^+e^-$ from the pair $\bar{t}t$ due to 
electromagnetic interaction, $<v>$ and $n$ as, respectively, a mean quark 
velocity and concentration of quarks (antiquarks) in toponium. To obtain 
$\sigma$ in the explicit form one may take the corresponding formula for the 
cross section of creation of $e^+e^-$ from the muon pair $\mu^+\mu^-$ 
(see, e. g., Ref. \cite{{LL1}}) and, after replacing 
$\alpha_{em}\to Q\alpha_{em}$, $m_{\mu}\to m_t$ with electromagnetic coupling 
constant $\alpha_{em}$=1/137.0359895 and muon mass $m_{\mu}$, obtain 
$$\sigma= \frac{4\pi N Q^2\alpha_{em}^2}{3s}\left(1+\frac{2m_e^2}{s}\right)
\sqrt{1-\frac{4m_e^2}{s}}\>,
\eqno(20) $$
where electron mass $m_e=0.510998918$ MeV, the Mandelstam invariant 
$s=2m_t(m_t+\epsilon/2)$ with $\epsilon$ from (12), $N$ is the number of 
colours and $Q=2/3$ for toponium. 
To get $<v>$ one may use 
the standard relativistic relation $v=\sqrt{T(T+2E_0)}/(T+E_0)$ with kinetic 
$T$ and rest energies $E_0$ for velocity $v$ of a point-like particle. Putting 
$T=\epsilon/2-m_t$, $E_0=m_t$ we shall gain 
$$<v>=\sqrt{1-\frac{4m_t^2}{\epsilon^2}} \>.  \eqno(21)$$
At last, obviously, $n=1/V$, where the volume of quarkonium $V=4\pi<r>^3/3$ 
with the sought $<r>$, the latter being yet not related to formula (16). 
The relations (19)--(21) 
entail the sought independent estimate for $<r>$

$$<r>=\left(\frac{3\sigma\sqrt{2}\sqrt{1-\frac{4m_t^2}{\epsilon^2}}}
{4\pi\Gamma_1}\right)^{1/3}                  \eqno(22)$$
with $\sigma$ of (20). When inserting $N=3, \epsilon=347.4$ GeV, $m_t= 173.25$ 
GeV, $m_e=0.510998918$ MeV, $\Gamma_1=13$ keV into (22) we shall have 
$<r>\approx0.2162653913\times 10^{-2}$ fm. In further considerations we can 
use this independent estimate of $<r>$ while calculating $<r>$ according to 
(16) which will impose certain restrictions on parameters of the confining 
SU(3)-gluonic field in toponium. 

It should be noted that in the heavy quarkonia physics (see, e. g., 
Ref. \cite{Bra05}) in (19) one often puts $n=|\psi(0)|^2$, where 
$\psi({\bf r})$ is a wave function of the heavy quarkonium stationary state 
which may be obtained, for example, within the framework of potential approach 
as a solution of the Schr{\"o}dinger type equation. Following this prescription 
in our approach with wave functions of (6) we should put 
$n=\sum\limits_{j=1}^3|\psi_j(0)|^2$ with $\psi_j$ of (6) which would entail 
$<r>=[3/(4\pi\sum\limits_{j=1}^3|\psi_j(0)|^2)]^{1/3}$ instead of (16). But it 
is clear that (16) gives physically more correct expression for $<r>$ since it 
employes all values of meson wave function rather than the only one at 
${\bf r}=0$ (inasmuch as here $<r>=\sqrt{\int r^2\psi^{\dag}\psi d^3x}=
\sqrt{\sum\limits_{j=1}^3\int r^2|\psi_j|^2d^3x}$). So we shall use 
just (16) in what follows. 

\section{Estimates for parameters of SU(3)-gluonic field in the ground state 
of toponium $\eta_t$}
Now we are able to estimate parameters $a_j, b_j, B_j$ of the confining 
SU(3)-field (3) for the toponium ground state $\eta_t$ within framework of two 
approaches -- relativistic and nonrelativistic ones. 

\subsection{Relativistic approach}
Under this approach we should consider (12) and (16) the system of equations 
which should be solved compatibly if taking $\epsilon= 347.4$ GeV, 
$m_t= 173.25$ GeV and $<r>\approx0.2162653913\times 10^{-2}$ fm in accordance 
with the independent estimate of Section 5. While computing 
for distinctness we take all eigenvalues $\lambda_j$ of the Euclidean Dirac 
operator ${\cal D}_0$ on the unit two-sphere ${\Bbb S}^2$ equal to (-1). The 
results of numerical compatible solving of equations (12), (16)
are adduced in Tables 1--2.

\begin{table}[htbp]
\caption{Gauge coupling constant, reduced mass $\mu_0$ and
parameters of the confining SU(3)-gluonic field in the toponium ground state 
$\eta_t$:  
relativistic approach}
\label{t.1}
\begin{center}
\begin{tabular}{|c|c|c|c|c|c|c|c|}
\hline
\small $ g$ & 
\small $\mu_0$ (\small GeV) & 
\small $a_1$ & 
\small $a_2$ & 
\small $b_1$ (\small GeV) & 
\small $b_2$ (\small GeV) & 
\small $B_1$ & \small $B_2$ \\
\hline
\scriptsize 1.24353  
& \scriptsize 86.6250 
& \scriptsize 0.361253
& \scriptsize 0.339442

& \scriptsize 48.9402
& \scriptsize 76.7974
& \scriptsize -0.360 
& \scriptsize  -0.295 \\
\hline
\hline
\end{tabular}
\end{center}
\end{table}

\begin{table}[htbp]
\caption{Theoretical ground state energy of toponium and its radius: 
relativistic approach}
\label{t.2}
\begin{center}
\begin{tabular}{|c|c|} 
\hline
\small Theoret. $\epsilon$ (GeV)  & 
\small Theoret. $<r>$ (fm)   \\
\hline
\small $\epsilon= 2m_t+
\omega_j(0,0,-1)= 347.400$  & 
\small $0.213915\times 10^{-2}$ \\
\hline
\end{tabular}
\end{center}
\end{table}

\subsection{Nonrelativistic approach}
If estimating the quark velocity in toponium by the relation (21) then 
at $\epsilon= 347.4$ GeV, $m_t= 173.25$ GeV we obtain 
$<v>\approx0.071935$ that points out a nonrelativistic approach to be 
applicable. It should be noted, however, this nonrelativistic approach should 
be consistent with the above relativistic one. We cannot, therefore, follow 
the standard strategy of the heavy quarkonia physics which exploits the 
so-called potential approach (see, e. g., Refs. \cite{{top},{Bra05}}). 
The essence of the latter is in that the interaction between quarks is modelled 
on a nonrelativistic confining potential in the form $V(R)=a/R+kR+c_0$ with some 
real constants $a, k, c_0$ and the distance between quarks $R$. 
However, parameters of such a potential, i.e. quantities $a,k,c_0$ are not 
related with QCD-Lagrangian in any way and we 
cannot speak about $V(R)$ as describing some gluon configuration between quarks. 
It would be possible if the mentioned potential were a solution of Yang-Mills 
equations directly derived from QCD-Lagrangian since, from the QCD-point of 
view, any gluonic field should be a solution of Yang-Mills equations (as well 
as any electromagnetic field is by definition always a solution of Maxwell 
equations). 

In reality, as was shown 
in Refs. \cite{{Gon051},{Gon052}} (see also Appendix $C$), potential of form 
$a/R+kR+c_0$ cannot be a solution of the Yang-Mills equations if 
simultaneously $a\ne0, k\ne0$. Therefore, 
it is impossible to obtain compatible solutions of the 
Yang-Mills-Dirac (Pauli, Schr{\"o}dinger) system when inserting potential of 
the mentioned form into Dirac (Pauli, Schr{\"o}dinger) equation. So, we draw 
the conclusion (mentioned as far back as in Refs. \cite{Gon03}) that the 
potential approach seems to be inconsistent: it is not based on compatible 
nonperturbative solutions for the Dirac-Yang-Mills system derived from 
QCD-Lagrangian in contrast to our confinement mechanism. Actually potential 
approach for heavy quarkonia has been historically modeled on positronium 
theory. In the latter case, however, one uses the {\em unique} modulo square 
integrable solutions of Dirac (Schr{\"o}dinger) 
equation in the Coulomb field [condensate of huge number of (virtual) photons], 
i. e., one employs the {\em unique} compatible nonperturbative solutions of the 
Maxwell-Dirac (Schr{\"o}dinger) system directly derived from QED-Lagrangian to 
describe positronium (or hydrogen atom) spectrum. 

On the other hand, as was mentioned in Section 1, our confinement mechanism is 
based on the {\em unique} family of compatible nonperturbative solutions for 
the Dirac-Yang-Mills system derived from QCD-Lagrangian and just magnetic 
colour field of solution (3) is 
responsible for linear confinement. But, as we have seen in Section 2 [see relation 
(10)], if directly taking the nonrelativistic limit $c\to\infty$ then 
the contribution of linear magnetic colour field (parameters $b_j, B_j$) to 
spectrum really vanishes and spectrum in essence becomes purely nonrelativistic 
Coulomb one (modulo the rest energy). Consequently, as we emphasized as far back as in 
Refs. \cite{Gon03}, the confinement mechanism under discussion is essentially 
connected with relativistic effects conditioned by availability of the 
mentioned magnetic colour field between any two quarks. Under the circumstances 
the only reasonable way of constructing a nonrelativistic approach within our 
confinement scheme is the power series expansion of the physical magnitudes 
of interest in $z=1/c$ with retaining necessary number of terms. It is clear, 
we then shall obtain the consistent transition from relativistic regime to 
nonrelativistic one. In their turn, the mentioned magnitudes may be computed 
within the relativistic framework with the help of wave functions (6) and then 
the necessary expansions should be fulfilled. 

Following the just formulated receipt for description of toponium in a nonrelativistic 
manner we should use (10) (at $\hbar=c=1$) with subtracting the rest energy 
$\mu_0c^2$ to replace the relations (12) (at $n_j=0, \lambda_j=-1$) by  
$$ \omega=-\mu_0g^2a_1^2/2-g^2a_1b_1+\mu_0g^3B_1a_1^2=  
-\mu_0g^2a_2^2/2-g^2a_2b_2+\mu_0g^3B_2a_2^2= $$
$$-\mu_0g^2a_3^2/2-g^2a_3b_3+\mu_0g^3B_3a_3^2=\epsilon-2m_t  \eqno(23)$$
with $a_3=-(a_1+a_2)$, $b_3=-(b_1+b_2)$, $B_3=-(B_1+B_2)$.  

At the same time we can use (16) to compute $<r>$ with replacing the 
quantities $\alpha_j=\sqrt{(\lambda_j-gB_j)^2-g^2a_j^2}$, 
$\beta_j=\sqrt{\mu_0^2-\omega_j^2+g^2b_j^2}$ by their nonrelativistic 
expressions 
$$ \alpha_j=|\lambda_j|-\frac{|\lambda_j|}{\lambda_j\hbar}gB_jz+ O(z^2)
\>,\eqno(24)$$
$$\beta_j= \mu_0c^2+\frac{1}{2\mu_0}
\left(g^2b_j^2-\frac{1}{4\hbar^4}\mu_0^2g^4a_j^4\right)z^2+O(z^3)\>. 
\eqno(25) $$
Before expanding $\beta_j$ we made replacement 
$\omega_j\to\omega_j-\mu_0c^2$
in the formula $\beta_j=\sqrt{\mu_0^2-\omega_j^2+g^2b_j^2}$, i. e. we 
subtracted the rest energy from $\omega_j$ as is required in nonrelativistic
limit (see, e. g., Ref. \cite{LL1}). After it we should compatibly solve 
equations (23) and (16) with $\alpha_j, \beta_j$ of (24)--(25) 
(with $\hbar=c=1$) at 
$\lambda_j=-1$, $\epsilon= 347.4$ GeV, $m_t= 173.25$ GeV and  
$<r>\approx0.2162653913\times 10^{-2}$ fm in accordance 
with the independent estimate of Section 5. When solving we should impose 
the conditions $\alpha_j>-1/2$, $\beta_j>0$ to have 
$\Psi_j\in L_2^{4}({\Bbb R}^3)$ at any 
$t\in{\Bbb R}$ for $\Psi_j$ of (6).

 The results of numerical computation are adduced in Tables 3--4. 

\begin{table}[htbp]
\caption{Gauge coupling constant, reduced mass $\mu_0$ and
parameters of the confining SU(3)-gluonic field in the toponium ground state 
$\eta_t$: nonrelativistic approach}
\label{t.3}
\begin{center}
\begin{tabular}{|c|c|c|c|c|c|c|c|}
\hline
\small $ g$ & \small $\mu_0$ (\small GeV) & \small $a_1$ 
& \small $a_2$ & \small $b_1$ (\small GeV) & \small $b_2$ (\small GeV) 
& \small $B_1$ & \small $B_2$ \\
\hline
\scriptsize 1.24353  
& \scriptsize 86.6250 
& \scriptsize 0.200 
& \scriptsize 1.32600
& \scriptsize -291.248
& \scriptsize 143.349
& \scriptsize -0.780 
& \scriptsize  1.40873 \\
\hline
\hline
\end{tabular}
\end{center}
\end{table}

\begin{table}[htbp]
\caption{Theoretical ground state energy of toponium and its radius: 
nonrelativistic approach}
\label{t.4}
\begin{center}
\begin{tabular}{|c|c|} 
\hline
\small Theoret. $\epsilon$ (GeV)  & 
\small Theoret. $<r>$ (fm)   \\
\hline
\small $\epsilon= 2m_t+
\omega_j(0,0,-1)= 347.400$  & 
\small $0.216264\times 10^{-2}$ \\
\hline
\end{tabular}
\end{center}
\end{table}

\section{Consistency with possible width of two-photon decay $\eta_t\to2\gamma$}

Let us consider whether the estimates of previous section are consistent with 
possible width $\Gamma_2$ of two-photon decay $\eta_t\to2\gamma$ 
which might be one of the main signals for detecting toponium under certain 
conditions (see Section 2). To estimate $\Gamma_2$ we can use an analogy of 
toponium with charmonium where width $\Gamma(\eta_c\to2\gamma)\approx7.2$ keV 
for the 
charmonium ground state $\eta_c$ whereas leptonic width 
$\Gamma(J/\psi\to e^+e^-)\approx5.55$ keV for state $J/\psi$ \cite{pdg}. 
Accordingly, ratio $\Gamma(\eta_c\to2\gamma)/\Gamma(J/\psi\to e^+e^-)
\approx1.297$ and if 
taking the same ratio for $\Gamma_2/\Gamma_1$ with $\Gamma_1\approx13$ keV of 
Section 5 for toponium state $V$ analogous to $J/\psi$ then we 
shall obtain $\Gamma_2\approx16.865$ keV. 

On the other hand, actually kinematic analysis based on 
Lorentz- and gauge invariances gives rise to the following expression for 
width $\Gamma$ of the electromagnetic decay $P\to2\gamma$ (where $P$ stands 
for any meson from 
$\pi^0$, $\eta$, $\eta^\prime$, see, e.g., Refs. \cite{RF})
$$ \Gamma=\frac{1}{4}\pi\alpha_{em}^2g^2_{P\gamma\gamma}\mu^3 \eqno(26) $$
with electromagnetic coupling constant $\alpha_{em}$=1/137.0359895 and 
$P$-meson mass $\mu$ while the information about strong 
interaction of quarks in $P$-meson is encoded in a decay constant 
$g_{P\gamma\gamma}$. Making replacement $g_{P\gamma\gamma}=
f_P/\mu $ we can reduce (26) to the form 
$$ \Gamma=\frac{\pi\alpha_{em}^2\mu f_P^2}{4}\>. \eqno(27) $$
Now it should be noted that the only 
invariant which $f_P$ might depend on is $Q^2=\mu^2$, i. e. we should find 
such a function ${\cal F}(Q^2)$ for that ${\cal F}(Q^2=\mu^2)=f_P$. It is 
obvious from physical point of view that ${\cal F}$ should be connected with 
electromagnetic properties of $P$-meson. As we have seen above in 
Section 4, there are at least two suitable functions for this aim -- electric 
and magnetic form factors. But there exist no experimental 
consequences related to magnetic form factor at present whereas electric 
one to some extent determines, e. g., an effective size of meson (quarkonium) 
in the 
form $<r>$ of (16). It is reasonable, therefore, to take 
${\cal F}(Q^2=\mu^2)=Af(Q^2=\mu^2)$ with some constant $A$ and electric form 
factor $f$ of (17) for the sought relation. Under the situation we obtain 
additional equation imposed on parameters of the confining SU(3)-gluonic field 
in $P$-meson which has been used in Refs. \cite{{Gon07a},{Gon07b}} for to 
estimate the mentioned parameters in $\pi^0$- and $\eta$-mesons. Inasmuch as 
relation (27) is in essence nonperturbative since decay constant $f_P$ cannot 
be computed by perturbative techniques, we may extend (27) over heavy quarkonia 
states similar to $\pi^0$, $\eta$, $\eta^\prime$, in particular, over $\eta_t$. 
As a result, using (17) we come from (27) to relation 
$$ \Gamma=\Gamma_2=\frac{\pi\alpha_{em}^2\mu}{4}
\left(A\sum\limits_{j=1}^3\frac{1}{6\alpha_jx_j}\cdot
\frac{\sin{(2\alpha_j\arctan{x_j})}}
{(1-x_j^2)^{\alpha_j}}\right)^2\approx16.865\,{\rm keV} \> \eqno(28) $$
with $x_j=\mu/(2\beta_j)$ and $\mu=\epsilon=347.4$ GeV. Under the circumstances 
we can employ the results of Tables 1 and 3 and compute the left-hand side of 
(28) in relativistic and nonrelativistic regimes respectively which entails 
the corresponding values $A\approx0.0340$ and $A\approx0.156$. Consequently, 
we draw the conclusion that parameters of the confining SU(3)-gluonic field 
in toponium from Tables 1 and 3 might be consistent with $\Gamma_2$ in both 
regimes. 
                                                            
\section{Estimates of gluon concentrations, electric and magnetic colour field 
strengths}
Now let us remind that, according to Refs. \cite{{Gon052},{Gon06}}, one can 
confront the field (3) with $T_{00}$-component (volumetric energy 
density of the SU(3)-gluonic field) of the energy-momentum tensor (2) so that 
$$T_{00}\equiv T_{tt}=\frac{E^2+H^2}{2}=\frac{1}{2}\left(\frac{a_1^2+
a_1a_2+a_2^2}{r^4}+\frac{b_1^2+b_1b_2+b_2^2}{r^2\sin^2{\vartheta}}\right)
\equiv\frac{{\cal A}}{r^4}+
\frac{{\cal B}}{r^2\sin^2{\vartheta}}\>\eqno(29)$$
with electric $E$ and magnetic $H$ colour field strengths and with real 
${\cal A}>0$, ${\cal B}>0$. One can also introduce magnetic colour induction 
$B=(4\pi\times10^{-7} {\rm H/m})\,H$, where $H$ in A/m.   

To estimate the gluon concentrations
we can employ (29) and, taking the quantity
$\omega= \Gamma_1=\Gamma_1(V\to e^+e^-)\approx13$ keV of Section 5 for 
the characteristic frequency of gluons, we obtain
the sought characteristic concentration $n$ in the form
$$n=\frac{T_{00}}{\Gamma_1}\> \eqno(30)$$
so we can rewrite (29) in the form 
$T_{00}=T_{00}^{\rm coul}+T_{00}^{\rm lin}$ conforming to the contributions 
from the Coulomb and linear parts of the
solution (3). This entails the corresponding split of $n$ from (30) as 
$n=n_{\rm coul} + n_{\rm lin}$. 

The parameters of Tables 1 and 3 were employed when computing and for 
simplicity we put $\sin{\vartheta}=1$ in (29), whereas the Bohr radius 
$a_0=0.529177249\cdot10^{5}\ {\rm fm}$ \cite{pdg}. 

Table 5 contains the numerical results for $n_{\rm coul}$, $n_{\rm lin}$, $n$, 
$E$, $H$, $B$ for the quarkonium under discussion in relativistic approach 
while Table 6 is obtained in nonrelativistic one.  
\begin{table}[htbp]
\caption{Gluon concentrations, electric and magnetic colour field strengths in 
toponium: relativistic approach}
\label{t.5}
\begin{center}
\begin{tabular}{|lllllll|}
\hline
\scriptsize $\eta_{t}$: & \scriptsize 
$r_0=<r>= 0.213915\times 10^{-2} \ {\rm fm}$ & & &  & & \\
\hline 
\tiny $r$ (fm)& \tiny $n_{\rm coul}$ $({\rm m}^{-3})$ & 
\tiny $n_{\rm lin}$ $({\rm m}^{-3})$  
& \tiny $n$ $({\rm m}^{-3})$ & \tiny $E$ $({\rm V/m})$ & 
\tiny $H$ $({\rm A/m})$ &\tiny $B$ $({\rm T})$\\
\hline
\tiny $0.1r_0$ 
& \tiny $ 0.880566\times10^{66}$   
& \tiny $ 0.131833\times10^{64}$ 
& \tiny $ 0.881884\times10^{66}$ 
& \tiny $ 0.223572\times10^{30}$  
& \tiny $ 0.116367\times10^{27}$ 
&  \tiny $0.146231\times10^{21}$\\
\hline
\tiny$r_0$ 
& \tiny$ 0.880566\times10^{62}$ 
& \tiny$ 0.131833\times10^{62}$ 
& \tiny$ 0.101240\times10^{63}$
& \tiny$ 0.223572\times10^{28}$  
& \tiny$ 0.116367\times10^{26}$  
&\tiny$  0.146231\times10^{20}$\\
\hline
\tiny$10r_0$ 
& \tiny$ 0.880566\times10^{58}$  
& \tiny$ 0.131833\times10^{60}$ 
& \tiny$ 0.140639\times10^{60}$ 
& \tiny$ 0.223572\times10^{26}$  
& \tiny$ 0.116367\times10^{25}$  
& \tiny$ 0.146231\times10^{19}$\\
\hline
\tiny$1.0$ 
& \tiny$ 0.184386\times10^{52}$  
& \tiny$ 0.603264\times10^{56}$ 
& \tiny$ 0.603282\times10^{56}$ 
& \tiny$ 0.102305\times10^{23}$  
& \tiny$ 0.248927\times10^{23}$                                   
& \tiny$ 0.312811\times10^{17}$  \\
\hline
\tiny$a_0$ 
& \tiny$ 0.235138\times10^{33}$  
& \tiny$ 0.215429\times10^{47}$ 
& \tiny$ 0.215429\times10^{47}$ 
& \tiny$ 0.365339\times10^{13}$ 
& \tiny$ 0.470404\times10^{18}$  
& \tiny$ 0.591127\times10^{12}$ \\
\hline
\end{tabular}
\end{center}
\end{table}

\begin{table}[htbp]
\caption{Gluon concentrations, electric and magnetic colour field strengths in 
toponium: nonrelativistic approach}
\label{t.6}
\begin{center}
\begin{tabular}{|lllllll|}
\hline
\scriptsize $\eta_{t}$: & \scriptsize 
$r_0=<r>= 0.216264\times 10^{-2} \ {\rm fm}$ & & &  & & \\
\hline 
\tiny $r$ (fm)
& \tiny $n_{\rm coul}$ $({\rm m}^{-3})$ 
& \tiny $n_{\rm lin}$ $({\rm m}^{-3})$  
& \tiny $n$ $({\rm m}^{-3})$ 
& \tiny $E$ $({\rm V/m})$ 
& \tiny $H$ $({\rm A/m})$ &\tiny $B$ $({\rm T})$\\
\hline
\tiny $0.1r_0$ 
& \tiny $ 0.402970\times10^{67}$   
& \tiny $ 0.680959\times10^{64}$ 
& \tiny $ 0.403651\times10^{67}$ 
& \tiny $ 0.478269\times10^{30}$  
& \tiny $ 0.264472\times10^{27}$ 
& \tiny $ 0.332345\times10^{21}$\\
\hline
\tiny$r_0$ 
& \tiny$ 0.402970\times10^{63}$ 
& \tiny$ 0.680960\times10^{62}$ 
& \tiny$ 0.471066\times10^{63}$
& \tiny$ 0.478269\times10^{28}$  
& \tiny$ 0.264472\times10^{26}$  
& \tiny$ 0.332345\times10^{20}$\\
\hline
\tiny$10r_0$ 
& \tiny$ 0.402970\times10^{59}$  
& \tiny$ 0.680960\times10^{60}$ 
& \tiny$ 0.721257\times10^{60}$ 
& \tiny$ 0.478269\times10^{26}$  
& \tiny$ 0.264472\times10^{25}$  
& \tiny$ 0.332345\times10^{19}$\\
\hline
\tiny$1.0$ 
& \tiny$ 0.881475\times10^{52}$  
& \tiny$ 0.318486\times10^{57}$ 
& \tiny$ 0.318494\times10^{57}$ 
& \tiny$ 0.223687\times10^{23}$  
& \tiny$ 0.571957\times10^{23}$ 
& \tiny$ 0.718742\times10^{17}$  \\
\hline
\tiny$a_0$ 
& \tiny$ 0.112410\times10^{34}$  
& \tiny$ 0.113733\times10^{48}$ 
& \tiny$ 0.113733\times10^{48}$ 
& \tiny$ 0.798800\times10^{13}$ 
& \tiny$ 0.108084\times10^{19}$  
& \tiny$ 0.135823\times10^{13}$ \\
\hline
\end{tabular}
\end{center}
\end{table}

\section{Discussion and concluding remarks}
\subsection{Discussion}
 We can see that main reasonable characteristics of $\eta_t$ such as energy 
of ground state and the root-mean-square radius (in essence, radius of confinement) 
and also width of possible decay $\eta_t\to2\gamma$ 
may be consistent with appropriate parameters of the confining SU(3)-gluonic 
field between quarks in toponium in both relativistic and nonrelativistic 
approaches within the framework of our confinement mechanism. In other words, 
we can obtain a description of toponium directly 
appealing to quark and gluonic degrees of freedom as should be from the first 
principles of QCD. As is seen from Tables 5 and 6, at both relativistic and 
nonrelativistic description the gluon concentrations are huge at the 
characteristic scales of the toponium ground state $\eta_t$ and the 
corresponding fields (electric and magnetic colour ones) can be considered 
to be the classical ones with enormous strengths. The part 
$n_{\rm coul}$ of gluon concentration $n$ connected with the Coulomb electric 
colour field is decreasing faster than $n_{\rm lin}$, the part of $n$ related 
to the linear magnetic colour field, and at large distances $n_{\rm lin}$ 
becomes dominant. It should be emphasized that in fact the gluon concentrations 
are much greater than the estimates given in Tables 5 and 6 because the latter 
are the estimates for sufficiently big possible gluon frequencies, i.e. for 
sufficiently big possible gluon impulses (under the 
concrete situation of toponium ground state $\eta_t$). As was mentioned in Section 1, 
the overwhelming majority of gluons between quarks should be soft, i. e., with 
frequencies $<<$ 13 keV so the corresponding concentrations $>>$ the ones in 
Tables 5 and 6. The given picture is in concordance with the one obtained 
in Refs. \cite{{Gon03},{Gon06},{Gon07a},{Gon07b}}. 
As a result, the confinement mechanism developed in 
Refs. \cite{{Gon01},{Gon051},{Gon052}} is also confirmed by the considerations 
of the present paper. 

By the way, the estimate of $<r>\approx0.216\times 10^{-2}$ fm obtained in 
present paper allows one to make a suggestion about why observation of toponium 
finds difficulty at colliders. Let us use an analogy with classical 
electrodynamics where, as is well known (see e. g. Ref. \cite{LL}), the notion 
of classical electromagnetic field (a photon condensate) generated by a charged 
particle is applicable only at distances $>>$ the Compton wavelength 
$\lambda_c=1/m$ for the given point-like particle with mass $m$. Passing on to 
QCD, gluons and quarkonia and replacing electromagnetic field by colour one 
in the case of $t$-quark with mass $m_t=173.25$ GeV we obtain 
$\lambda_c\approx0.11\times10^{-2}$ fm. If comparing the above $<r>$ to 
characteristic radius of weak interaction 
$r_{weak}\sim1/m_W\approx0.2450\times10^{-2}$ fm 
($m_W\approx80.403$ Gev is mass of $W$-boson) and to the just obtained 
$\lambda_c$ then we have inequality $\lambda_c<r_{weak}\sim\,<r>$ so one 
may draw the conclusion that in pair $\bar{t}t$ when creating at colliders the 
most probable distance between quarks is $\le r_{weak}$ so that quarks are 
more inclined to weak interaction rather than to strong one. In other words, 
they have not time in order to form a classical confining SU(3)-gluonic field 
due to strong interaction and to constitute a bound state in virtue of it. But 
we cannot completely exclude events where distance between $t$-quarks when creating 
at colliders would be much greater than $\lambda_c$ which might entail 
formation of $\eta_t$, for example, and, consequently, a signal for detecting 
toponium, e.g., in the form of decay $\eta_t\to2\gamma$, as mentioned in 
Section 2. 

It should be noted, however, that our results are of a preliminary character 
not only because of that the experimental spectroscopy of toponium is in its 
infancy but also in virtue of that the current quark masses 
(as well as the gauge coupling constant $g$) used in computation are known 
only within the certain limits and we can expect similar limits for the 
magnitudes discussed in the paper so it is necessary further specification of 
the parameters for the confining SU(3)-gluonic field 
in toponium which can be obtained when experimental situation for toponium 
becomes more satisfactory. We hope to then continue analysing the toponium 
physics.  
\subsection{Concluding remarks}
Finally we should note the following. As has been shown in the paper, our 
approach allows one to conduct both relativistic and nonrelativistic 
description and both the cases are consistent with each other. Only 
experiments can, however, determine what physical picture (relativistic or 
nonrelativistic one) for quarkonia 
is really realized and enough for their complete description. Our approach 
works in either case since it is based on the unique family of compatible 
nonperturbative solutions for the Dirac-Yang-Mills system derived from 
QCD-Lagrangian and, as a result, the approach is itself nonperturbative, 
relativistic from the outset, admits self-consistent nonrelativistic limit 
and may be employed for any meson (quarkonium). 


\section*{Appendix A}
We here represent some results about eigenspinors of the Euclidean Dirac 
operator on two-sphere ${\Bbb S}^2$ employed in the main part of the paper. 

When separating variables in the Dirac equation (4) there naturally 
arises the Euclidean Dirac operator ${\cal D}_0$ on the unit two-dimensional 
sphere ${\Bbb S}^2$ and we should know its eigenvalues with the corresponding 
eigenspinors. Such a problem also arises in the black hole theory while 
describing the so-called twisted spinors on Schwarzschild and 
Reissner-Nordstr\"om black holes and it was analysed in 
Refs. \cite{{Gon052},{Gon99}}, so we can use the results obtained 
therein for our aims. Let us adduce the necessary relations. 

The eigenvalue equation for
corresponding spinors $\Phi$ may look as follows
$${\cal D}_0\Phi=\lambda\Phi.\>\eqno(A.1)$$

As was discussed in Refs. \cite{Gon99}, the natural form of ${\cal D}_0$ in 
local coordinates $\vartheta, \varphi$ on the unit sphere ${\Bbb S}^2$ looks 
as 
$${\cal D}_0=-i\sigma_1\left[
i\sigma_2\partial_\vartheta+i\sigma_3\frac{1}{\sin{\vartheta}}
\left(\partial_\varphi-\frac{1}{2}\sigma_2\sigma_3\cos{\vartheta}
\right)\right]=$$
$$\sigma_1\sigma_2\partial_\vartheta+\frac{1}{\sin\vartheta}
\sigma_1\sigma_3\partial_\varphi- \frac{\cot\vartheta}{2}
\sigma_1\sigma_2         \eqno(A.2)$$
with the ordinary Pauli matrices
$$\sigma_1=\pmatrix{0&1\cr 1&0\cr}\,,\sigma_2=\pmatrix{0&-i\cr i&0\cr}\,,
\sigma_3=\pmatrix{1&0\cr 0&-1\cr}\,, $$
so that $\sigma_1{\cal D}_0=-{\cal D}_0\sigma_1$.

The equation $(A.1)$ was explored in Refs. \cite{Gon99}.
Spectrum of $D_0$ consists of the numbers
$\lambda=\pm(l+1)$              
with multiplicity $2(l+1)$ of each one, where $l=0,1,2,...$. Let us 
introduce the number $m$ such that $-l\le m\le l+1$ and the corresponding 
number $m'=m-1/2$ so $|m'|\le l+1/2$. Then the conforming eigenspinors of  
operator ${\cal D}_0$ are 
$$\Phi=\pmatrix{\Phi_1\cr\Phi_2\cr}= 
\Phi_{\mp\lambda}=\frac{C}{2}\pmatrix{P^k_{m'-1/2}\pm P^k_{m'1/2}\cr
P^k_{m'-1/2}\mp P^k_{m'1/2}\cr}e^{-im'\varphi}\> \eqno(A.3) $$
with the coefficient $C=\sqrt{\frac{l+1}{2\pi}}$ and $k=l+1/2$. 
These spinors form an orthonormal basis in $L_2^2({\Bbb S}^2)$ 
and are subject 
to the normalization condition
$$\int_{{\Bbb S}^2}\Phi^{\dag}\Phi d\Omega=
\int\limits_0^\pi\,\int\limits_0^{2\pi}(|\Phi_{1}|^2+|\Phi_{2}|^2)
\sin\vartheta d\vartheta d\varphi=1\>. \eqno(A.4)$$
Further, owing to the relation $\sigma_1{\cal D}_0=-{\cal D}_0\sigma_1$ we, 
obviously, have
$$ \sigma_1\Phi_{\mp\lambda}=\Phi_{\pm\lambda}\,.  \eqno(A.5)$$

As to functions $P^k_{m'n'}(\cos\vartheta)\equiv P^k_{m',\,n'}(\cos\vartheta)$ 
then they can be chosen by 
miscellaneous ways, for instance, as follows (see, e. g.,
Ref. \cite{Vil91})
$$P^k_{m'n'}(\cos\vartheta)=i^{-m'-n'}
\sqrt{\frac{(k-m')!(k-n')!}{(k+m')!(k+n')!}}
\left(\frac{1+\cos{\vartheta}}{1-\cos{\vartheta}}\right)^{\frac{m'+n'}{2}}\,
\times$$
$$\times\sum\limits_{j={\rm{max}}(m',n')}^k
\frac{(k+j)!i^{2j}}{(k-j)!(j-m')!(j-n')!}
\left(\frac{1-\cos{\vartheta}}{2}\right)^j \eqno(A.6)$$
with the orthogonality relation at $m',n'$ fixed
$$\int\limits_0^\pi\,{P^{*k}_{m'n'}}(\cos\vartheta)
P^{k'}_{m'n'}(\cos\vartheta)
\sin\vartheta d\vartheta={2\over2k+1}\delta_{kk'}
\>.\eqno(A.7)$$
It should be noted that square of 
${\cal D}_0$ is 
$${\cal D}^2_0=-\Delta_{{\Bbb S}^2}I_2+
\sigma_2\sigma_3\frac{\cos{\vartheta}}{\sin^2{\vartheta}}\partial_\varphi
+\frac{1}{4\sin^2{\vartheta}} +\frac{1}{4}\>,
\eqno(A.8)$$
while Laplacian on the unit sphere is
$$\Delta_{{\Bbb S}^2}=
\frac{1}{\sin{\vartheta}}\partial_\vartheta\sin{\vartheta}\partial_\vartheta+
\frac{1}{\sin^2{\vartheta}}\partial^2_\varphi=
\partial^2_\vartheta+\cot{\vartheta}\partial_\vartheta
+\frac{1}{\sin^2{\vartheta}}\partial^2_\varphi\>,
\eqno(A.9)$$
so the relation $(A.8)$ is a particular case of the so-called 
Weitzenb{\"o}ck-Lichnerowicz formulas (see Refs. \cite{81}). 
Then from $(A.1)$ it follows 
${\cal D}^2_0\Phi=\lambda^2\Phi$ and, when using the ansatz  
$\Phi=P(\vartheta)e^{-im'\varphi}=\pmatrix{P_1\cr P_2\cr}e^{-im'\varphi}$, 
$P_{1,2}=P_{1,2}(\vartheta)$, the equation ${\cal D}^2_0\Phi=\lambda^2\Phi$ 
turns into 
$$\left(-\partial^2_\vartheta-\cot{\vartheta}\partial_\vartheta +
\frac{m'^2+\frac{1}{4}}{\sin^2{\vartheta}}+
\frac{m'\cos{\vartheta}}{\sin^2{\vartheta}}\sigma_1\right)P=$$
$$\left(\lambda^2-\frac{1}{4}\right)P\>,
\eqno(A.10)$$
wherefrom all the above results concerning spectrum of ${\cal D}_0$ can be 
derived \cite{Gon99}.

When calculating the functions $P^k_{m'n'}(\cos\vartheta)$ directly, to our 
mind, it is the most convenient to use the integral expression \cite{Vil91}

$$P^k_{m'n'}(\cos\vartheta)=\frac{1}{2\pi}
\sqrt{\frac{(k-m')!(k+m')!}{(k-n')!(k+n')!}}\>
\int_{0}^{2\pi}\left(e^{i\varphi/2}\cos{\frac{\vartheta}{2}}+
ie^{-i\varphi/2}\sin{\frac{\vartheta}{2}}\right)^{k-n'}\times$$
$$\left(ie^{i\varphi/2}\sin{\frac{\vartheta}{2}}+
e^{-i\varphi/2}\cos{\frac{\vartheta}{2}}\right)^{k+n'}e^{im'\varphi}d\varphi 
\eqno(A.11)$$
and the symmetry relations ($z=\cos{\vartheta}$) 
$$P^k_{m'n'}(z)=P^k_{n'm'}(z), \>P^k_{m',-n'}(z)=P^k_{-m',\,n'}(z), 
\>P^k_{m'n'}(z)=P^k_{-m',-n'}(z)\,,$$ 
$$P^k_{m'n'}(-z)=i^{2k-2m'-2n'}P^k_{m',-n'}(z)\>. \eqno(A.12)$$
In particular
$$P^{k}_{kk}(z)=
\cos^{2k}{(\vartheta/2)},  
P^{k}_{k,-k}(z)=i^{2k}\sin^{2k}{(\vartheta/2)},
P^{k}_{k0}(z)=\frac{i^{k}\sqrt{(2k)!}}{2^k k!}\sin^{k}{\vartheta}\,,$$
$$ P^{k}_{kn'}(z)=i^{k-n'}\sqrt{\frac{(2k)!}{(k-n')!(k+n')!}}
\sin^{k-n'}{(\vartheta/2)}\cos^{k+n'}{(\vartheta/2)}\>. \eqno(A.13)$$ 

If $\lambda=\pm(l+1)=\pm1$ then $l=0$ and from $(A.3)$ it follows that 
$k=l+1/2=1/2$, $|m'|<1/2$ and we need the functions $P^{1/2}_{\pm1/2,\pm1/2}$ 
that are easily evaluated with the help of $(A.11)$--$(A.13)$ so   
the eigenspinors for $\lambda=-1$ are 
$$\Phi=\frac{C}{2}\pmatrix{\cos{\frac{\vartheta}{2}}+
i\sin{\frac{\vartheta}{2}}\cr
\cos{\frac{\vartheta}{2}}-i\sin{\frac{\vartheta}{2}}\cr}e^{i\varphi/2}, 
\Phi=\frac{C}{2}\pmatrix{\cos{\frac{\vartheta}{2}}+
i\sin{\frac{\vartheta}{2}}\cr
-\cos{\frac{\vartheta}{2}}+i\sin{\frac{\vartheta}{2}}\cr}
e^{-i\varphi/2},\eqno(A.14)$$
while for $\lambda=1$ the conforming spinors are
$$\Phi=\frac{C}{2}\pmatrix{\cos{\frac{\vartheta}{2}}-
i\sin{\frac{\vartheta}{2}}\cr
\cos{\frac{\vartheta}{2}}+i\sin{\frac{\vartheta}{2}}\cr}e^{i\varphi/2}, 
\Phi=\frac{C}{2}\pmatrix{-\cos{\frac{\vartheta}{2}}+
i\sin{\frac{\vartheta}{2}}\cr
\cos{\frac{\vartheta}{2}}+i\sin{\frac{\vartheta}{2}}\cr}e^{-i\varphi/2}
\eqno(A.15) $$
with the coefficient $C=\sqrt{1/(2\pi)}$.

It is clear that $(A.14)$--$(A.15)$ can be rewritten in the form 

$$\lambda=-1: \Phi=\frac{C}{2}\pmatrix{e^{i\frac{\vartheta}{2}}
\cr e^{-i\frac{\vartheta}{2}}\cr}e^{i\varphi/2},\> {\rm or}\>\>
\Phi=\frac{C}{2}\pmatrix{e^{i\frac{\vartheta}{2}}\cr
-e^{-i\frac{\vartheta}{2}}\cr}e^{-i\varphi/2},$$
$$\lambda=1: \Phi=\frac{C}{2}\pmatrix{e^{-i\frac{\vartheta}{2}}\cr
e^{i\frac{\vartheta}{2}}\cr}e^{i\varphi/2}, \> {\rm or}\>\>
\Phi=\frac{C}{2}\pmatrix{-e^{-i\frac{\vartheta}{2}}\cr
e^{i\frac{\vartheta}{2}}\cr}e^{-i\varphi/2}\,, 
\eqno(A.16) $$
so the relation $(A.5)$ is easily verified at $\lambda=\pm1$. 

\section*{Appendix B}
We here adduce the explicit form for the radial parts of meson wave functions 
from (6). At $n_j=0$ they are given by 
$$F_{j1}=C_jP_jr^{\alpha_j}e^{-\beta_jr}\left(1-
\frac{Y_j}{Z_j}\right),F_{j2}=iC_jQ_jr^{\alpha_j}e^{-\beta_jr}\left(1+
\frac{Y_j}{Z_j}\right),\eqno(B.1)$$
while at $n_j>0$ by
$$F_{j1}=C_jP_jr^{\alpha_j}e^{-\beta_jr}\left[\left(1-
\frac{Y_j}{Z_j}\right)L^{2\alpha_j}_{n_j}(r_j)+
\frac{P_jQ_j}{Z_j}r_jL^{2\alpha_j+1}_{n_j-1}(r_j)\right],$$
$$F_{j2}=iC_jQ_jr^{\alpha_j}e^{-\beta_jr}\left[\left(1+
\frac{Y_j}{Z_j}\right)L^{2\alpha_j}_{n_j}(r_j)-
\frac{P_jQ_j}{Z_j}r_jL^{2\alpha_j+1}_{n_j-1}(r_j)\right]\eqno(B.2)$$
with the Laguerre polynomials $L^\rho_{n}(r_j)$, $r_j=2\beta_jr$, 
$\beta_j=\sqrt{\mu_0^2-\omega_j^2+g^2b_j^2}$ at $j=1,2,3$ with 
$b_3=-(b_1+b_2)$, 
$P_j=gb_j+\beta_j$, $Q_j=\mu_0-\omega_j$,
$Y_j=P_jQ_j\alpha_j+(P^2_j-Q^2_j)ga_j/2$, 
$Z_j=P_jQ_j\Lambda_j+(P^2_j+Q^2_j)ga_j/2$    
with $a_3=-(a_1+a_2)$,   
$\Lambda_j=\lambda_j-gB_j$ with $B_3=-(B_1+B_2)$, 
$\alpha_j=\sqrt{\Lambda_j^2-g^2a_j^2}$, 
while $\lambda_j=\pm(l_j+1)$ are
the eigenvalues of Euclidean Dirac operator ${\cal D}_0$ 
on unit two-sphere with $l_j=0,1,2,...$ (see Appendix A) 
and quantum numbers $n_j=0,1,2,...$ are defined by the relations 
$$n_j=\frac{gb_jZ_j-\beta_jY_j}{\beta_jP_jQ_j}\,, 
\eqno(B.3)$$
which entails the quadratic equation (7) and spectrum (8).  
Further, $C_j$ of $(B.1)$--$(B.2)$ should be determined
from the normalization condition
$$\int_0^\infty(|F_{j1}|^2+|F_{j2}|^2)dr=\frac{1}{3}\>.\eqno(B.4)$$
As a consequence, we shall gain that in (6) 
$\Psi_j\in L_2^{4}({\Bbb R}^3)$ at any $t\in{\Bbb R}$ and, accordingly,
$\Psi=(\Psi_1,\Psi_2,\Psi_3)$ may describe relativistic bound states 
in the field (3) with the energy spectrum (8). As is clear from $(B.3)$ at 
$n_j=0$ we have 
$gb_j/\beta_j=Y_j/Z_j$ so the radial parts of $(B.1)$ can be rewritten as  

$$F_{j1}=C_jP_jr^{\alpha_j}e^{-\beta_jr}\left(1-
\frac{gb_j}{\beta_j}\right),F_{j2}=iC_jQ_jr^{\alpha_j}e^{-\beta_jr}\left(1+
\frac{gb_j}{\beta_j}\right)\>.\eqno(B.5)$$

More details can be found in Refs. \cite{{Gon01},{Gon052}}. 
\section*{Appendix C}
 The facts adduced here have been obtained in 
Refs. \cite{{Gon01},{Gon051},{Gon052}} and we concisely give them here only 
for completeness of discussion in Section 6. 

The Dirac-Yang-Mills system derived from QCD-Lagrangian  
according to the standard prescription of Lagrange approach is 
$${\cal D}\Psi=\mu_0\Psi\>,\eqno(C.1)$$
$$d\ast F= g(\ast F\wedge A - A\wedge\ast F) +gJ\>\eqno(C.2)$$
with a gauge coupling constant $g$, Dirac operator ${\cal D}$, 
$F=dA+gA\wedge A$ and the 
Cartan's wedge (external) product $\wedge$, whereas $\ast$ means the Hodge 
star operator conforming to a Minkowski metric, for instance, in the form of 
(1), while the source $J$ (a non-Abelian SU($3$)-current) is 
$$J=j_\mu^a\lambda_a\ast(dx^\mu)=\ast j=\ast(j_\mu^a\lambda_adx^\mu)=
\ast(j^a\lambda_a)\>,\eqno(C.3)$$
where currents 
$$j^a=j_\mu^adx^\mu=\overline{\Psi}(I_3\otimes\gamma_\mu)\lambda^a\Psi\,dx^\mu\>,$$
so summing over 
$a=1,...,8$ is implied in ($C.3$). Besides we have 
${\rm div}(j^a)={\rm div}(j)=0$ if 
$\Psi$ obeys Dirac equation $C.1$ \cite{{Gon01},{Gon052}}, 
where the divergence of the Lie algebra valued
1-form $A=A_\mu dx^\mu=A^a_\mu \lambda_adx^\mu$ is defined by the relation 
(see, e. g. Refs. \cite{81})
$${\rm div}(A)=\frac{1}{\sqrt{\delta}}\partial_\mu(\sqrt{\delta}g^{\mu\nu}
A_\nu)\>.$$ 
Definitions of the operators $\ast$ and $d$ (external differentiation) can 
be found in Refs. \cite{81} while explicit form of Dirac operator ${\cal D}$ 
of $(C.1)$ depends on choice of 
local coordinates on Minkowski spacetime and for case of coordinates 
$t, r, \vartheta, \varphi$ Dirac equation $(C.1)$ can be rewritten in 
form (4)--(5) of Section 3 and if we require its modulo square 
integrable solutions to consist from the components 
of form $\Psi_j\sim r^{\alpha_j}e^{-\beta_j r}$ with some $\alpha_j>0$, 
$\beta_j>0$ then it will entail all the components of the current $J$ to be 
modulo $<< 1$ at each point of Minkowski space (perhaps except for a small 
neighbourhood of point ${\bf r}=0$). The latter allow us to put $J\approx0$ 
and we come to the problem of finding the confining solutions for the 
Yang-Mills equations of $(C.2)$ with $J=0$ whose unique nontrivial form is 
given by (3) of Section 3 while the unique corresponding modulo square 
integrable solutions of $C.1$ are given by (6) of Section 3 (for more details 
see Refs. \cite{{Gon01},{Gon051},{Gon052}} and Appendix B). 

As has been mentioned in Subsection 6.2, in meson spectroscopy one often uses
the nonrelativistic confining 
potentials. Those confining potentials between quarks 
are usually modelled in the form $a/r+br$ with some constants $a$ and $b$.
It is clear, however, that from the QCD point of view the interaction between
quarks should be described by the whole SU(3)-field $A_\mu=A^a_\mu \lambda_a$,
genuinely relativistic object, the nonrelativistic potential being only some
component of $A^a_t$ surviving in the nonrelativistic limit when the light 
velocity $c\to\infty$. Let us explore whether such potentials may be the 
solutions of the Maxwell or SU(3)-Yang-Mills equations. Though this can be 
easily derived from the above result about uniqueness of solution (3) of 
Section 3, let us consider the given situation directly in view of its 
physical importance. We shall use the Hodge star operator action on the 
basis differential 1- and 2-forms on Minkowski spacetime with local 
coordinates $t, r, \vartheta, \varphi$ in the form 
$$\ast dt=r^2\sin{\vartheta}dr\wedge d\vartheta\wedge d\varphi,\>
\ast dr=r^2\sin{\vartheta}dt\wedge d\vartheta\wedge d\varphi,\>$$
$$\ast d\vartheta=-r\sin{\vartheta}dt\wedge dr\wedge d\varphi,\>
\ast d\varphi=rdt\wedge dr\wedge d\vartheta,\>$$
$$\ast(dt\wedge dr)=-r^2\sin\vartheta d\vartheta\wedge d\varphi\>,
\ast(dt\wedge d\vartheta)=\sin\vartheta dr\wedge d\varphi\>,$$
$$\ast(dt\wedge d\varphi)=-\frac{1}{\sin\vartheta}dr\wedge d\vartheta\>,
\ast(dr\wedge d\vartheta)=\sin\vartheta dt\wedge d\varphi\>,$$
$$\ast(dr\wedge d\varphi)=-\frac{1}{\sin\vartheta}dt\wedge d\vartheta\>.
\eqno(C.4)$$

\subsubsection*{Maxwell Equations}
      In the case of Maxwell equations (i.e., Yang-Mills equations for the case 
of U(1)-group looking as $d\ast F=0$ at $J=0$) the ansatz $A=
A_tdt=(a/r+br)dt$ yields $F=dA=(a/r^2-b)dt\wedge dr$. Then with the help 
of $(C.4)$ we have
$*F=\sin\vartheta(br^2-a)d\vartheta\wedge d\varphi$ and the relation
$d*F=2br\sin\vartheta dr\wedge d\vartheta\wedge d\varphi=0$ entails
$b\equiv0$. 
      
\subsubsection*{SU(3)-Yang-Mills Equations}
We use the ansatz 
$$A=A^a_t\lambda_adt=(A'/r+B'r)dt\>\eqno(C.5)$$ 
with some constant matrices $A'=\alpha^a\lambda_a, B'=\beta^a\lambda_a$. Then 
$A\wedge A=0$, $F=dA+gA\wedge A=dA=(A'/r^2-B')dt\wedge dr$. Again with the help 
of $(C.4)$ we have $\ast F=\sin{\vartheta}(B'r^2-A')d\vartheta\wedge d\varphi$, 
$d\ast F= 2B'r\sin{\vartheta}dr\wedge d\vartheta\wedge d\varphi$, 
$\ast F\wedge A-A\wedge\ast F=
-2[A',B']rdt\wedge d\vartheta\wedge d\varphi$. 
Under the circumstances the Yang-Mills equations ($C.2$) (at $J=0$) are 
tantamount to the conditions 
$d\ast F=0$, $\ast F\wedge A-A\wedge\ast F=0$. The former entails 
$B'=0$, then the latter is fulfilled at any $A'$ and we can see that the 
Coulomb-like field $A=(A'/r)dt$ is a solution of the 
Yang-Mills equations ($C.2$) (at $J=0$) with arbitrary constant matrix $A'$.   
In principle the ansatz $(C.5)$ might be a solution of ($C.2$) with the source 
of the form
$$J=2B'r\sin{\vartheta}dr\wedge d\vartheta\wedge d\varphi+
2g[A',B']rdt\wedge d\vartheta\wedge d\varphi=$$
$$\ast j=\ast(j^a_\mu\lambda_a dx^\mu)=\ast\left(
\frac{2B'}{r}dt+g\frac{2[A',B']}{r\sin{\vartheta}}dr\right)\>,\eqno(C.6)$$
but ${\rm div}(j)\ne0$ and this is not consistent with the only source $(C.3)$ 
derived from the QCD-Lagrangian. We can avoid this difficulty putting 
matrices $A', B'$ are not equal to zero simultaneously and both matrices 
belong to Cartan subalgebra of SU(3)-Lie algebra, i.e. commutator 
$[A',B']=0$. Then 
$B'=\beta_3\lambda_3+\beta_8\lambda_8$ and for consistency with the only 
admissible source of $(C.3)$ we should require source of $(C.3)$ to be equal 
to one of $(C.6)$ which entails 
$$g\overline{\Psi}(I_3\otimes\gamma_\mu)\lambda^a\Psi\lambda_a\,dx^\mu=
\frac{2(\beta^3\lambda_3+\beta^8\lambda_8)}{r}dt\>,$$ 
wherefrom one can conclude that 
$$g\overline{\Psi}(I_3\otimes\gamma_t)\lambda^a\Psi=0, a\ne3, 8\>,
g\overline{\Psi}(I_3\otimes\gamma_t)\lambda^3\Psi=\frac{2\beta^3}{r}\>,$$
$$ g\overline{\Psi}(I_3\otimes\gamma_t)\lambda^8\Psi=
\frac{2\beta^8}{r}\>,
g\overline{\Psi}(I_3\otimes\gamma_\mu)\lambda^a\Psi=0, a=1,...,8,\,
\mu\ne t\>,\eqno(C.7)$$
which can obviously be satisfied only at 
$\beta^3\sim\beta^8\sim \Psi\to 0$ at each point of Minkowski spacetime, i. e., 
really matrix $B'=0$ again. All the above can easily be generalized to 
any group SU($N$) with $N>1$ \cite{{Gon051},{Gon052}}.  

As a result, the potentials employed in nonrelativistic approaches do 
not obey the Maxwell or Yang-Mills equations.
The latter ones are essentially relativistic and, as we can see from 
(3) of Section 3, the components linear in $r$ of the unique exact solution 
$A_\mu$ are different from $A_t$ and related with magnetic (colour) field 
vanishing in the nonrelativistic limit.

\subsection*{Remark about search for nonrelativistic confining potentials} 
The above results make us cast a new glance at search of many years for 
nonrelativistic potentials modelling the confinement. Many efforts were 
devoted to the latter topic, for example, within the framework of lattice 
gauge theories or potential approach (see, e.g., Ref. \cite{Bra05} and 
references therein). It should be noted, however, that almost in all 
literature on this direction one does not bring up the question: whether such 
potentials could (or should) satisfy the Yang-Mills equations? As is clear 
from the above the answer is negative. That is why the mentioned approaches 
seem to be inconsistent: such potentials cannot describe any gluonic 
configuration between quarks since any gluonic field should be a solution of 
Yang-Mills equations (as well as any electromagnetic field is by definition 
always a solution of Maxwell equations).


\end{document}